\documentclass[amsmath,twocolumn,showpacs,aps,prl]{revtex4}
\usepackage{bm}
\usepackage{graphics}
\usepackage{graphicx}

\begin{document}

\title{Resistivity effects in surface superconductivity
of thin films in strong magnetic fields}
\author{A. A. Zyuzin and A. Yu. Zyuzin}

\affiliation{ A. F. Ioffe Physico-Technical Institute of Russian
Academy of Sciences, 194021 St. Petersburg, Russia}


\begin{abstract}
Phase slips creation in the thin film in perpendicular magnetic
filed with edge superconductivity is studied. These centers are
due to thermal activation of the order parameter below
superconducting temperature transition leading to the suppression
of the superconductivity. The corresponding resistance is
calculated. The Alsamazov- Larkin correction to the conductivity
above the critical magnetic field destroying the surface
superconductivity is studied. Such structures could be applied as
a new system for the study of the phase slip phenomenon in one-
dimensional superconducting wires.
\end{abstract}

\maketitle

As it was first shown by Saint- James and de Gennes \cite{deGennes}, superconductivity can nucleate in the thin surface
superconducting sheath to magnetic fields higher than the bulk
critical field $H_{c2} < H< H_{c3} \approx 1.69 H_{c2}$. 

Thin films reveal the most simple picture of surface superconductivity.
In particular, this case was studied experimentally in papers
\cite{MagneticImp, Scola}, where the temperature
dependencies of resistivity of thin $\mathrm{Nb}$ films were
measured. Effects of the surface inhomogeneities and the sample
shape, properties of mesoscopic size superconductors in surface
superconductivity regime were investigated both theoretically and
experimentally in papers \cite{Baelus, Dikin}.

However, it is of definite interest to consider the limit of the
extremely thin superconducting film, where superconductivity
persists in the quasi- one- dimensional edge layer.

It is well known that the fluctuations of the order parameter play
an important role in the physics of low- dimensional
superconductors (thin films, wires). At temperatures above the
critical temperature of the superconducting transition $T_{c}$
fluctuations lead to the enhancement of the conductivity \cite{AslamazovLarkin}, while below $T_{c}$ they destroy the long- range
order and lead to the finite resistance of the system.

In the vicinity of $T_{c}$ for example in thin superconducting
wires with the diameter smaller than the coherence length thermal
activation of the phase slips centers locally destroys the
superconductivity \cite{LA, MH, Lukens, Newbower}. Phase slip event is of the order of the coherence
length $\xi(T)$ where the amplitude of the order parameter
vanishes at one point while the phase difference between the
opposite sites of this point is $\pi$.

The observed resistance of the extremely thin superconducting
wires at temperatures $T<<T_{c}$ \cite{Giordano, Lau} is
argued to be caused by the phase slip events due to quantum
tunneling of the order parameter \cite{Zaikin}.

In the present paper we will show that similar fluctuation effects
appear in the edge superconductivity of thin superconducting film
in the magnetic field perpendicular to the plane of the film. To
the best of our knowledge the question of the fluctuations of the
order parameter in the vicinity of the phase transition in thin
films with edge superconductivity still remains open.

We will give the detailed analysis of the sample resistivity
dependencies on temperature and magnetic field. The equation
for Aslamazov- Larkin correction to the conductivity at fields
higher than $H_{c3}(T)$ will be also obtained.

\begin{figure}[t] \centering
\includegraphics[width=7.0cm]{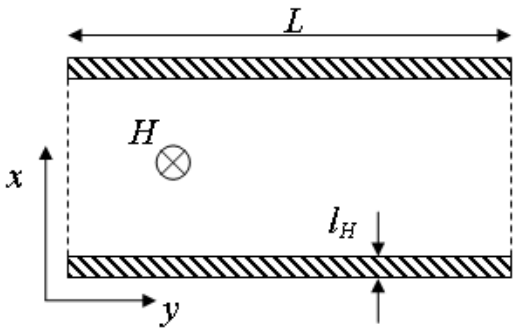} \caption{Thin film of thickness
$d$ in perpendicular magnetic field $H$, magnetic length $l_{H}$
is the width of the superconducting layer, $L$- the length of the
layer along the edge of the film} \label{fig1}
\end{figure}

Let us consider a thin superconducting film with the magnetic
field applied perpendicular to the surface of the film (see
Fig.\ref{fig1}). We will study the case of the 2-type
superconductor under the surface superconductivity condition, when
the Ginzburg- Landau parameter $\kappa \gg 1$. Thus, we will not
take into account the magnetic field modulations due to
supercurrents. The size of the film is such that $d \ll\xi (T)\ll
L$, where $d$ and $L$ is the width and length of the film. The
Ginzburg- Landau equation in dimensionless variables could be
written in the form
\begin{equation}\label{NLGL}
(i\mathbf{\nabla}+\mathbf{A})^2\Psi=\Psi(1-|\Psi|^2)
\end{equation}
where $\Psi$ is the complex order parameter, length is measured in
units of the coherence length $\xi(T)=(\hbar\pi D /
8(T_{c}(H)-T))^{1/2}$, $\mathbf{A}$- vector potential measured in
units $\frac{c\hbar}{2e\xi(T)}$, $D$- diffusion coefficient.

Since the superconductivity in this regime exists only in the thin
edge layer of the film we can treat each edge independently. Then
let $y$ axis to be applied along the corresponding edge, $x$ axis
directed to the bulk of the film. Magnetic filed is applied along the
$z$-axis. Vector potential is chosen in the Landau gauge,
$\mathbf{A}=(0,Hx,0)$. The boundary condition for the order
parameter at the $(0,y)$ edge of the film is given as
\begin{equation}\label{bd}
\frac{d\Psi}{dx}\mid_{x=0}=0
\end{equation}
at the same time, the order parameter vanishes at the bulk of the
film. We will search for the solution of the nonlinear Ginzburg-
Landau equation (\ref{NLGL}) in the form
\begin{equation}\label{g}
\Psi_{0}(x,y)=\gamma g(x)e^{iky}
\end{equation}
where $\gamma$ is some constant, $g(x)$ is a function subject to
the condition $\int_{0}^{\infty}g^{2}(x)dx =1$. Then the equation
(\ref{NLGL}) can be reformulated as
\begin{equation}
V(x)=|\gamma g_{0}(x)|^2, E_{0}=0
\end{equation}
where $g_{0}(x)$ is the eigenfunction corresponding to the lowest
eigenvalue $E_{0}$ of the equation
\begin{equation}\label{lin}
\left(-\frac{d^{2}}{dx^2}+\left(Hx-k\right)^2-1 +V(x)\right)g(x)=Eg(x)
\end{equation}

In order to solve equation (\ref{lin}) we can use the perturbation
theory for $\gamma \ll 1$. The eigenfunctions are then expressed
through the parabolic cylinder function \cite{Abrikos}
$g_{n}(x)\propto D_{a}\left( x\sqrt{2H}-k\sqrt{\frac{2}{H}}
\right)$, where $a=\frac{E_{n}-H+1}{2H}$.

The eigenvalue of the linear equation (\ref{lin}) corresponding to
the ground state of the system is a function of $k$ and has a
minimum where
\begin{equation}\label{E0}
E_{0}= \left[ k-k_{0}(H)\right]^2-\epsilon +\gamma^2
\int_{0}^{\infty}g_{0}^{4}(x)dx
\end{equation}
where $\epsilon= 1-H/H_{c3}$ and $k_{0}(H)\sim \sqrt{H}$. First
two terms in the equation (\ref{E0}) can be obtained from the
boundary condition $\frac{d}{dx}D_{a}\left(
x\sqrt{2H}-k\sqrt{\frac{2}{H}} \right)|_{x=0}$.

The ground state eigenvalue condition $\min E_{0}=0$ describes the
superconducting transition point. The choice of $k=k_{0}(H)$
corresponds to the zero current case, further we will treat only
this case. Finally, we obtain
\begin{equation}
\gamma^2=\epsilon \left(\int_{0}^{\infty} dxg_{0}^4(x)\right)^{-1}
\end{equation}

\begin{figure}[h] \centering
\includegraphics[width=8.0cm]{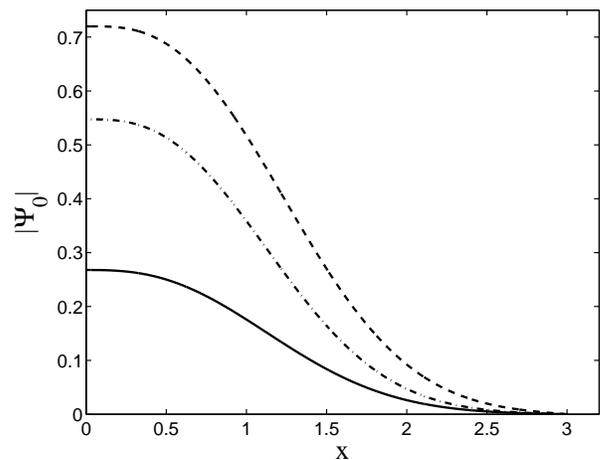} \caption{The
amplitude of the order parameter $|\Psi_{0}|$ for the $H=1$,
$k=0.73$ (dashed line); $H=1.3$, $k=0.85$ (dash- dot); $H=1.6$,
$k=0.93$ (solid line)} \label{fig2}
\end{figure}

We also provide the numerical solution for the nonlinear Ginzburg-
Landau equation. Fig. \ref{fig2} shows the zero current case
$k=k_{0}(H)$ solution for different values of the magnetic field.
Indeed, the order parameter is localized in the vicinity of the
edge of the film at distance of the order of the magnetic length.
The amplitude of the order parameter vanishes with increasing the
magnetic field. The deviations from the solution obtained by the
perturbation theory are small up to $H\approx H_{c2}$.

For practical purpose \cite{Abrikos} it is convenient to
approximate the solution for the $g(x)$ by the function
\begin{equation}\label{g2}
\tilde{g}(x)=\left(\frac{4bH}{\pi}\right)^{1/4}e^{-\frac{bHx^2}{2}}
\end{equation}
where $b=\sqrt{1-2/\pi}$ and $H_{c3} = 1/b \simeq 1.66$, that
differs from the exact one only by the 2\%, while
$k_{0}(H)=\sqrt{H/b\pi}$. Using equations (\ref{g}) and (\ref{g2})
we find the approximate solution to the nonlinear Ginzburg
-Landau equation
\begin{equation}\label{F1}
\tilde{\Psi}_{0}(x,y)=(\epsilon \sqrt{2})^{1/2}e^{-\frac{H
}{2H_{c3}}x^2}e^{ik_{0}y}
\end{equation}

It is well known that phase slip events in 1D superconducting wires \cite{LA}
as well as vortices creation in superfluid liquid \cite{Iordanski} at $T\leq T_{c}$ are due to thermal
activation of the order parameter. The order
parameter switches between the metastable states of the
superconductor changing phase by $2\pi$. The probability of such
process is governed by the Arrhenius law $\propto \exp{(-\Delta
F/T)}$, where $\Delta F$ is the energy barrier separating these
states. Solution for the phase slip event corresponds to the
saddle point of the barrier.

\begin{figure}[h] \centering
\includegraphics[width=8.0cm]{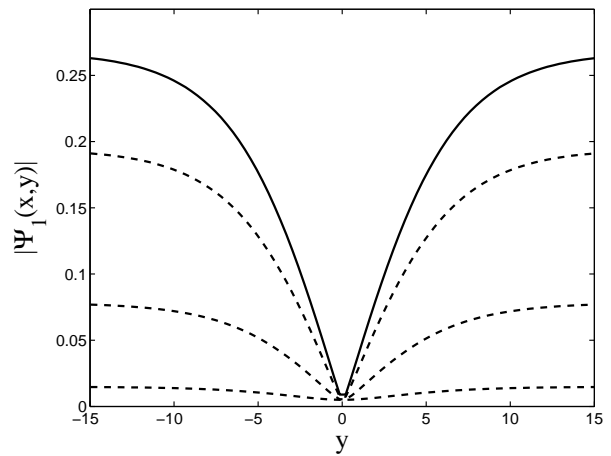} \caption{The amplitude
of the order parameter $|\Psi_{1}(x,y)|$ for the case $H=1.6$,
$k=0.93$; solid line corresponds to the solution on the boundary
$x=0$, dashed line corresponds to $x=1$, $x=1.6$, $x=2.3$}
\label{fig3}\end{figure}

We have found the numerical solution for the phase slip center in
the edge layer of the thin film by analyzing the time- dependent
Ginzburg- Landau equation (for the review see \cite{Tinkham})
with the periodic boundary conditions on the order parameter. The
solution for the amplitude of the order parameter is shown in Fig.
\ref{fig3} for different values of $x$ at the magnetic field
$H=1.6$, $k=0.93$, which corresponds to $\epsilon= 0.1$.
In order to derive the approximate analytical solution for the
phase slip center we will search for the order parameter in the
form $\Psi_{1}(x,y)= \gamma e^{ik_{0}y}\sum_{n\geq
0}C_{n}(y)g_{n}(x)$, where the summation is over the set of
eigenfunctions of the equation (\ref{lin}). In the zero-mode
regime taking into account only $n=0$, we find
\begin{equation}\label{1DGL}
\frac{d^2C}{dy^2}+\epsilon C(1-C^2)=0
\end{equation}
This equation is similar to the Ginzburg- Landau equation for the
one- dimensional wire at zero applied current case. Solving
equation (\ref{1DGL}) we obtain the expression for the order
parameter corresponding to phase slip event in surface
superconductivity
\begin{equation}\label{F2}
\tilde{\Psi}_{1}(x,y)=\tilde{\Psi}_{0}(x,y)\tanh\left(y\sqrt{\epsilon / 2}\right)
\end{equation}

Taking into account solution (\ref{F2}) we conclude according to
\cite{LA} with the expression for the resistivity of the thin
film superconducting edge layer.
\begin{equation}
R=\frac{\pi\hbar^{2} \Omega}{2e^{2}T} \exp\left(-\Delta F/T\right)
\end{equation}
where
\begin{equation}\Delta F=\frac{b\sqrt{2}H_{c3}^2(T)}{16\pi
\kappa^2}[\ell^2_{H_{c3}}d]\int dx dy \left(|\Psi_0|^4-|\Psi_1|^4
\right)
\end{equation}
is the saddle-point free energy barrier increment,
$\Omega=(L/\xi(T))(\epsilon^{3/2}/\tau_{GL})\left(\Delta F
/T\right)^{1/2}$ is the attempt frequency, $L$ is the length of
the edge superconducting layer,
$\tau_{GL}=[\pi\hbar/8(T_{c3}(H)-T)]$ is the relaxation time and
$\kappa$ is the GL parameter.

Using equations (\ref{F1}) and (\ref{F2}) at $H<H_{c3}(T)$ we find
\begin{equation}
\Delta F =
\frac{bH_{c3}^2(T)}{12\sqrt{\pi}\kappa^2}\left[\ell_{H_{c3}}\ell_{H}d\right]\epsilon^{3/2}
\end{equation}
where $d$- is the film thickness, $\ell_{H_{c3}}=\sqrt{\hbar
c/eH_{c3}(T)}$- is the magnetic length.

This expression is simply the condensation energy of the
superconductivity in volume $\ell_{H_{c3}}\ell_{H}d$ of thin film
superconducting edge layer.

Notice that the width of the edge superconducting layer $\ell_{H}$
at $T\rightarrow T_{c}(H)$ is much smaller than the length of the
normal part of the layer $\ell_{H_{c3}}/\sqrt{\epsilon}$ caused by
the thermal activation of the phase slip event, pointing the
applicability of the thin wire approximation to the surface
superconductivity of thin film.

Phase slip events in 1D superconducting wires at low temperatures
are argued to be due to quantum tunneling \cite{Giordano, Lau, Zaikin}. The resistivity of the wire is then $R\propto
\exp{(-2S)}$, where the exponent of the tunneling amplitude is
$S=A\frac{\hbar}{e^2}G_{\xi}$ and $A$ is a constant of the order
of unity, $G_{\xi}$ is the conductance of the wire of length
$\xi(T)$ \cite{Zaikin}. In our case of the edge
superconducting layer it is the magnetic length $\ell_{H_{c3}}$
that governs the tunneling amplitude.
The tunneling process is accompanied by the creation of the
acoustic plasmons \cite{Shon} which are responsible for the
interaction between the phase slip centers. As a result, this
interaction suppresses the tunneling probability and this effect
is stronger as smaller the plasmons velocity, i.e. as stronger the
Coulomb screening. The dissipation effects are the second factor
that leads to decreasing of the probability of the phase slip centers
due to quantum tunneling \cite{Buch}.

In contrast to the case of isolated superconducting wire, the
Coulomb interactions in the regime of edge superconductivity is
screened by the charge of the normal part of the film. This
immediately leads to decreasing of the acoustic plasmons velocity.
The interaction with the normal part of the film results in the
effective relaxation of the order parameter phase fluctuations.
Consequently, the resistivity of the surface superconducting layer
of the thin film should be lower than the resistivity of the wire,
taken under the same conditions.

Let us consider the fluctuations of the order parameter in the
case of surface superconductivity at magnetic fields
$H>H_{c3}(T)$. The corresponding Aslamazov- Larkin correction to
the conductivity at magnetic fields higher than the
superconducting transition  field $H_{c3}$ was studied in the
paper \cite{Schmidt}. It was shown that for the case of two-
dimensional surface superconducting layer this correction has the
same temperature dependence as for the thin film at $T>T_{c}$.

However, in present work we focus on the extreme case of surface
superconductivity, when the order parameter is concentrated in the
quasi-one- dimensional layer of the thin film.

Fluctuations of the order parameter at magnetic fields $H>
H_{c3}(T)$ result in additional correction to the conductance
which according to \cite{Tinkham} is
\begin{equation}
G =\frac{(2e)^2}{2m}\sum_{\nu}\langle |\phi_{\nu}|^2 \rangle
\frac{\tau_{\nu}}{2}
\end{equation}
where summation goes over the set $\nu= n,k$. The value of
fluctuation of the order parameter
$\Psi(x,y)=\sum_{\nu}\phi_{\nu}g_{\nu}(x)e^{iky}$ is written as
\begin{equation}
\langle |\phi_{\nu}|^2 \rangle= \left[\frac{2m
\ell^{2}_{H_{c3}}}{b\hbar^2}\right]\frac{T}{E_{n}(k)}
\end{equation}
while $\tau_{\nu}=\tau_{GL}/E_{n}(k)$ is the characteristic decay
time of the fluctuation. The lowest eigenvalue with $n=0$ gives
the main contribution to the sum and coming from summation to the
integration over $k$ we obtain the correction to the conductance
of unit length of the layer
\begin{equation}\label{popravka}
G \simeq 0.28 \frac{e^2}{\hbar} \frac{H_{c3}(0)}{H_{c3}(T)}
\frac{\ell_{H_{c3}}}{|\epsilon|^{3/2}}
\end{equation}

Notice that the functional dependence of the correction $G \propto
|\epsilon|^{-3/2}$ is similar to the case of one- dimensional
superconducting wire.

It is seen also, with decreasing the temperature $H_{c3}(T)$
increases and the value of the Aslamazov- Larkin correction
decreases. However, at the same time equation (\ref{popravka}) still
valid in the interval of the magnetic fields $H- H_{c3}(T)$
that increases with decreasing the temperature.

For the numerical estimations we take typical values $H_{c3}\sim 1
$T then $\ell_{H_{c3}}\sim 25$nm. For ultrathin film $d \sim
10$nm, $\kappa =10$, we estimate at $T_{c3} \sim 1$K for the value
$\Delta F/T \sim 10^{3}\epsilon^{3/2}$. The probability of the
phase slip event becomes negligibly small unless the parameter
$\epsilon=1-H/H_{c3}$ is of the order of $ 10^{-2}$.

According \cite{Abrikos} the critical current destroying the
surface superconductivity could be written as $ J_{c}\simeq j_{c}
\ell_{H_{c3}} d$, where $j_{c}=\frac{1}{3\pi
\sqrt{6}}\frac{cH_{c}(T)}{\kappa \xi(T)}$ is the critical current
density of thin wire, we estimate $J_{c} \sim 20$$\mu A$.

To summarize, we have shown that the phase slip phenomenon reveals
in edge superconducting layer of thin film in perpendicular
magnetic field at $H<H_{c3}(T)$. The corresponding resistance was
calculated. The Aslamazov- Larkin correction to the edge
superconductivity of thin film at $H>H_{c3}(T)$ have also been
obtained. We conclude that such structures could be applied as a
new system for the study of the phase slip phenomenon in one-
dimensional superconducting wires.

We thank V.I. Kozub for helpful discussions and valuable
questions. The research was supported by Dynasty foundation, INTAS
Grant 05- 109- 4829 and RFFI Grant 06- 02- 17047.

\end{document}